\documentclass{article}

\usepackage[square,comma,sort&compress]{natbib}

\usepackage{dcolumn}
\newcolumntype {d} [1] {D{.}{.}{#1}}            

\usepackage{multirow}

\usepackage{threeparttable}

\usepackage{bigstrut}

\usepackage{tabularx}

\usepackage{exscale}

\usepackage{textcomp}

\usepackage{bm}

\usepackage{paralist}

\usepackage{xspace}

\usepackage{graphicx}

\newcommand \dif     {\ensuremath{\mathrm{d}}}	 		                              
\newcommand \pd  [2] {\ensuremath{\frac{\partial #1}{\partial #2}}}         
\newcommand \td  [2] {\ensuremath{\frac{\dif #1}{\dif #2}}}                 

\newcommand {\kBT} {\ensuremath{k_{\mathrm{B}}T}\xspace}              
\newcommand {\pH} {\ensuremath{\mathrm{pH}}\xspace}                   
\newcommand {\um} {\mbox{\textmu m}\xspace}                           
\newcommand {\un} [1] {\ensuremath{\,\mathrm{#1}}}          

\newcommand {\ie} {\emph{i.e.}~}						
\newcommand {\iec} {\emph{i.e.,}\xspace}		

\newcommand \eqref [1] {\mbox{eq.~(\ref{#1})}}		
\newcommand \figref [1] {\mbox{fig.~\ref{#1}}}		
\DeclareGraphicsExtensions{.eps}      

\title{Pulling adsorbed polymers from surfaces with the AFM: 
stick versus slip, peeling versus gliding}

\author{A.~Serr and R.~R.~Netz \thanks{corresponding author: netz@ph.tum.de}\\
Physics Department, TU Munich --- 85748 Garching, Germany}

\begin{document}

\maketitle

\begin{abstract}
We consider the response of an adsorbed polymer that is pulled by an AFM
within a simple geometric framework. We separately consider
the cases of i) fixed polymer-surface contact point, ii) sticky case where the 
polymer is peeled off from the substrate, and iii) slippery case where the
polymer glides over the surface. The resultant behavior depends on the value
of the surface friction coefficient and the adsorption strength. Our
resultant force profiles in principle allow to extract both from non-equilibrium
force-spectroscopic data.
62.25.+g (Mechanical properties of nanoscale materials)\\
82.35.Gh (Polymers on surfaces; adhesion)
\end{abstract}

After its introduction in the 1980s \cite{Binnig86}, the atomic-force microscope 
(AFM) has been intensely used to study the mechanical properties of 
single molecules \cite{Moy94}. Applications range from sequential unfolding 
of collapsed biopolymers over stretching of coiled synthetic polymers to breaking individual 
covalent bonds \cite{Rief97,Ortiz99,Grandbois99, Janshoff00a}. 
In recent experiments, the desorption of polyelectrolytes such as DNA, poly-(vinylamines), and poly-(acrylicacid) and polymers in varying solvent conditions physisorbed to different substrates was investigated \cite{Chatellier98,Hugel01,Haupt02,Friedsam04a}. 

Depending on the adsorption strength between polymer and substrate,
AFM single-polymer studies split in two classes: In the first, 
the applied forces are relatively weak so that 
the attachment on the cantilever tip and on the substrate is irreversible 
up to a certain maximal force and over the typical experimental time-scales; 
in this case the measured distance-force
traces contain information on the polymer that is being stretched and can 
be used to extract the polymer elasticity by comparison with molecular models \cite{Kreuzer01,Hugel05}.
In the second class, the applied force is strong enough to detach
the polymer from the substrate. 
In this case, the measured force-distance relation 
contains information about the strength of the surface-polymer interaction
and, as we will show in this paper, about the nanoscopic friction 
effects at the substrate. 
In fact, assuming that the polymer glides very easily over the surface and 
surface friction can be neglected,
plateau forces are measured the 
heights of which correspond to the adsorption free energy per unit length \cite{Chatellier98,Hugel01,Hanke05}.
In the presence of finite surface-polymer friction, 
the force-distance curves exhibit more complex behavior.

\begin{figure}[hb]
	\includegraphics[width=144mm]{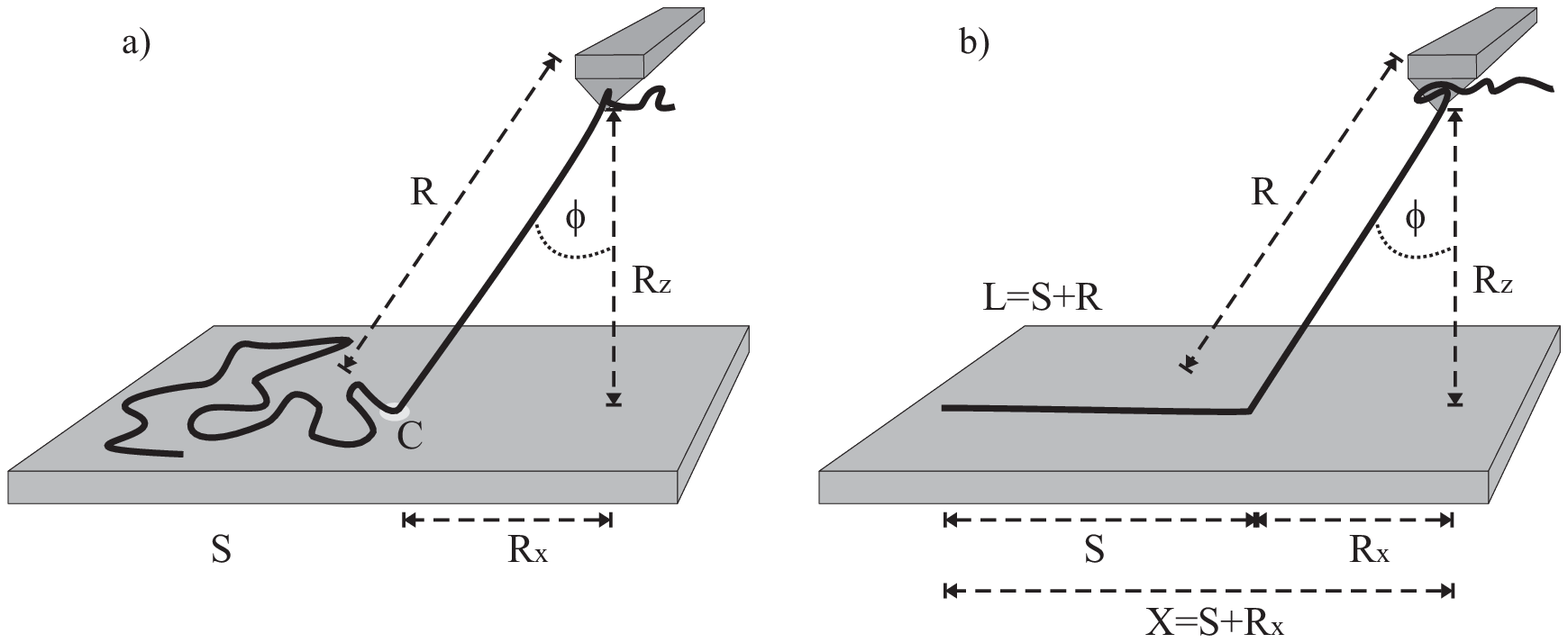}
  \caption {Schematic geometry of a single-polymer experiment
with the AFM. The polymer is bound between a planar surface and the cantilever tip. 
a)~Tilted geometry with the adsorbed polymer strand
exhibiting a self-similar, crumpled conformation, b)~Idealized geometry where the
adsorbed polymer strand is linearly stretched and aligned in parallel with
the detached polymer section.}
	\label{fig:def-var}
\end{figure}

In the interpretation of force-distance curves it is often assumed that the polymer is vertically attached between cantilever tip and surface. This is not necessarily true, and in this
paper we point out the consequences of a non-zero attachment angle $\phi$ as defined schematically in \figref{fig:def-var}.
In the case of irreversible attachment between
polymer and substrate, \ie where the polymer-substrate contact point 
is immobile, the angle $\phi$ is fixed for a given vertical distance between the polymer-substrate 
and polymer-cantilever contact points and determined by their lateral distance.
This distance is typically not controlled in experiments,
but the resulting force-distance curve decisively depends on this angle.
In the case of reversible attachment between polymer and substrate, 
the resultant behavior depends crucially on the surface friction 
coefficient of the polymer: For very large friction coefficient, the polymer
is peeled off from the substrate segment by segment but does not slide; 
here the angle is dictated by geometric considerations and
changes as the polymer is peeled from the surface. 
Force-distance curves in this case depend sensitively on the lateral surface configuration of the polymer.
For small friction coefficient (or, equivalently, for small pulling rates), on the other hand,
the polymer portion in contact with the surface is sliding over the surface
and the angle adjusts according to a balance of friction and 
adsorption forces at the contact point. All these geometric
considerations have a bearing on the force-distance curves.
Proper analysis of the non-equilibrium features of the force curves
allows to extract friction coefficients of single polymers on surfaces
and thus an important parameter characterizing the nano-tribology
of adsorbed polymers.

\paragraph{Fixed contact point\\}
\label{sec:fixed-contact-point}
We first consider a polymer attached to the surface at some fixed contact
point along the polymer contour length. Experimentally, this can be achieved by 
covalent bonding between reactive surface groups and polymer monomers \cite{Hugel05}.
The schematic geometry is given in \figref{fig:def-var}a, where
the extension of the polymer stretched between the contact point $C$ and
the cantilever tip is denoted by $R$, and its lateral and vertical components
by $R_x$ and $R_z$, respectively. Clearly, 
  $R^2 = R_x^2 + R_z^2$.
We denote the elastic free energy of the polymer in units of \kBT by $W(R)$, which
contains entropic as well as energetic contributions due to deformation of bonds. 
We neglect the coupling to the probing device, as is justified
for sufficiently stiff cantilevers or optical traps \cite{Liphardt02,Kreuzer01a}.
In the presence of an external force $F_z$ acting along the vertical direction,
 the total free energy in units of \kBT becomes
\begin{equation}
  \label{eq:engery-with-pulling}
  E = W(R) - R_z F_z / (\kBT) = W(\sqrt{R_x^2 + R_z^2}) - R_z F_z / (\kBT) \,.
\end{equation}
The equilibrium extension of the polymer follows by minimization
of the free energy, 
  $\partial E/\partial R_z=0$,
leading to
\begin{equation}
  \label{eq:pulling-force}
  \frac{F_z b}{\kBT} = \frac{W'(R_z \alpha ) b}{\alpha }\,,
\end{equation}
where
  $W'(x) = \partial W(x) / \partial x$ 
is the derivative of the elastic polymer energy 
and $b$ the bond or Kuhn length.
As a measure of the chain orientation
we define the geometric factor
\begin{equation}
  \label{eq:geo-fac}
  \alpha = 1/\cos \phi = \sqrt{1 + R_x^2 / R_z^2} =  (1 -  R_x^2 / R^2)^{-1/2} \,.
\end{equation}
For perfect vertical alignment one has $\alpha =1$, for a slanted chain one finds $\alpha > 1$. 
For small stretching forces, any polymer behaves like a harmonic entropic spring 
with spring constant
  $K = 3 \kBT /(2 R_0^2)$ 
where $R_0^2$ is the mean-squared
end-to-end radius of the unperturbed chain. The elastic free energy reads
  $W(x)=K x^2 /2$ and thus $W'(x)=K x$.
In this case, the geometric factors in 
\eqref{eq:pulling-force} exactly cancel and the force-distance relation 
becomes independent of the chain orientation,
  ${F_z b/(\kBT) = b K R_z}$.
This cancellation only occurs for a harmonic elastic free energy;
in general, a non-trivial dependence arises. 
For the case of a freely jointed chain, characterized by a bond length $b$ 
and contour length $R_L$, the
elastic force at large stretching reads
  $W'(x) = (1/b)(1 - x / R_L)^{-1}$.
Insertion into \eqref{eq:pulling-force} yields \cite{Livadaru03}
  $b \alpha F_z / (\kBT) = ( 1 - \alpha  R_z / R_L )^{-1} $.
For a worm-like chain characterized by a persistence length
  $\ell \approx b/2$,
the  force in the large-stretching limit reads \cite{Livadaru03} 
  $W'(x) = 1/(4 \ell )(1 - x / R_L)^{-2}$.
The force-distance relation becomes 
  $4 \ell \alpha F_z / (\kBT) = ( 1 - \alpha  R_z / R_L )^{-2} $.
In both cases, the geometric factor $\alpha$ can be interpreted as renormalizing
the bond length $b$ (or persistance length $\ell$) 
and the chain contour length $R_L$. 
To get an estimate for the typical values of $\alpha$, we assume the chain
to perform a random walk on the surface prior to pick-up by the cantilever,
characterized by a swelling exponent $\nu$. 
Further assuming that the chain extension $R$ approximately equals
the contour length $R_L$ of the stretched segment, $R \approx R_L$, we can write
  $R_x \simeq b(R/b)^\nu$
and thus obtain
  $\alpha =  (1 -  (b/R)^{2-2\nu} )^{-1/2}  \approx 1 + (b/R)^{2-2\nu} / 2$.
For a self-avoiding walk on the surface, one has $\nu=3/4$ and thus
  $\alpha \approx 1 + (b/R)^{1/2} / 2$.
Thus, for a ds-DNA chain with Kuhn length $b=100\un{nm}$ and total length $R=10\,\um$, 
the geometric factor evaluates to $\alpha \approx 1.05$, and
thus leads to a 5 per cent variation in fitted values for persistence length and chain length. 
The effect drops with increasing polymer length and decreasing Kuhn length and is in fact
negligible in many practical cases.
On the other hand, the renormalized force-distance relation eq. (2) can 
be directly checked by AFM experiments with lateral position resolution. 

\paragraph{Sticky case\\}
\label{sec:sticky}
We now turn to the case where the adsorption of the polymer on the substrate
is reversible and thus the contact point can move via de- or adsorption.
For simplicity, we assume that the adsorption energy per
Kuhn length $b$ which is given by $\omega b$ satisfies
  $\omega b \gg 1$, 
where $\omega$ is the adsorption energy per $\kBT$ and unit length.
This implies that we are in a strong adsorption regime
and the polymer forms a flat quasi-two-dimensional layer on the surface \cite{Netz96}.
The total free energy of the adsorbed polymer strand of contour length $S$ is
\begin{equation}
  \label{eq:adsorbed-energy}
  E = - \omega S \,.
\end{equation}
In the following, we neglect elastic deformations of the desorbed polymer strand 
which is assumed to be fully stretched to its contour length, \ie $R=R_L$, thus preventing monomer-monomer contacts,
and disregard any temperature dependence of the adsorption strength
$\omega$ which is treated as a phenomenological parameter\cite{Orlandini,Mishra}.
We first consider infinite friction of the polymer at the surface: the polymer
will thus stick on the surface and a sufficiently strong force will peel the polymer off
from the surface. As the contact point moves over the surface, what is the resultant vertical force
on the cantilever?
The initial geometry is specified by arbitrary values $S_0$, $R_0$, $R_{z0}$, and $R_{x0}$.
We define the polymer contour length that has been peeled off as
  $P \equiv S_0 - S = R - R_0$, 
and parameterize all other variables by $P$. 
Assuming again that the adsorbed polymer
shows a self-similar lateral distribution function, we find 
\begin{equation}
  \label{eq:lateral}
	R_x^2 (P) \simeq R^2_{x0} + b^2 (P/b)^{2 \nu} \,.
\end{equation}
The vertical force acting on the cantilever tip can be calculated from
 \eqref{eq:adsorbed-energy} as
\begin{equation}
  \label{eq:sticky-peeling-force}
	\frac{F_z}{\kBT} = \pd{E}{R_z} = \pd{E}{P} \pd{P}{R_z} = \omega \pd{P}{R_z} \,,
\end{equation}
where we used
  $S = S_0 - P$. 
From \eqref{eq:lateral} and
  $R^2 (P) = R_x^2 (P) + R_z^2 (P)$ it follows that
\begin{equation}
  \label{eq:sticky-factor}
	\pd{R_z(P)}{P} = 
	  \frac{ R_0 + P - \nu b (P/b)^{2\nu - 1} } {R_z} =
	 \frac{ R_0 + P - \nu b (P/b)^{2\nu - 1} } 
	 {\sqrt{(R_0 + P)^2 - R_{x0}^2 - b^2 (P/b)^{2 \nu} }} \,.
\end{equation}
For a crumpled polymer, characterized by $ \nu < 1$, and for large peeling length 
  $P \rightarrow \infty$, 
the above relation crosses over to 
  $\partial R_z / \partial P \simeq 1$. 
The vertical measured force thus reaches a finite plateau value
  $F_z/(\kBT) = \omega$.
On the contrary, for a polymer which is adsorbed straight on the substrate, characterized by 
$\nu = 1$, one has
  $\partial R_z (P)/\partial P=R_0 / R_z$,
and thus
\begin{equation}
  \label{eq:force-straightened-stick}
	F_z / (\kBT) = \omega R_z / R_0
\end{equation}
implying that in this case the force increases linearly with the vertical distance. 

\paragraph{Slippery case\\}
\label{sec:slip}
We now assume a finite polymer-surface friction coefficient so that
sliding of the polymer on the substrate is possible when the cantilever
is moved either vertically or horizontally. 
On the cantilever tip the polymer is supposed to stick. 
When the polymer follows the cantilever motion and glides over 
the surface, friction forces lead to partial alignment;
we therefore simplify the discussion by assuming the polymer to be completely
stretched on the surface, as shown in \figref{fig:def-var}b.
We define $S$ and $R$ as the contour lengths of the adsorbed and desorbed polymer parts,
while
  $L=S+R$ 
is the total contour length.
The end-point position of the polymer relative to the tip is denoted as
  $X = S + R_x$.
The geometry is fully determined by two length scales, we choose as 
parameters the end-point position $X$ and the cantilever height $R_z$.
The total friction force is proportional to the sliding velocity, 
  $\dot{X} = d X / d t$,
and the length of the adsorbed part $S$, and acts 
parallel to the sliding direction,
\begin{equation}
  \label{eq:fric}
	\frac {F_{x}^{fric}} {\kBT} = \dot{X} S \zeta = \dot{X} \frac {L^2 - X^2 - R_z^2} {2 (L - X)} \zeta \,.
\end{equation}
Here $\zeta$ is the sliding friction coefficient per unit length and \kBT. 
It depends on all polymer and surface characteristics and is
in addition influenced by $\pH$, ionic strength, etc.
In this work it is assumed independent of the adsorption strength $\omega$ and
the pulling velocity.
The friction force is balanced by the horizontal component of the adsorption force, which we
associate with the spatial derivative of the adsorption energy \eqref{eq:adsorbed-energy}. 
We neglect any dependence of the adsorption energy on the gliding velocity
and note that \eqref{eq:adsorbed-energy} in the present non-equilibrium context
is not a free energy but rather corresponds to the non-dissipative contribution
to the work of desorption. 
It follows that
\begin{equation}
  \label{eq:ads}
	\frac {F_{x}^{ads}} {\kBT}= - \left(\pd {E}{X} \right)_{R_z} = \frac{\omega}{2} \frac{L^2 + X^2 - 2L X - R_z^2} {(L - X)^2 } = \omega \frac {R_x} {R_x - R} \,.
\end{equation}
Equating friction and adsorption forces yields a differential equation for the polymer geometry
\begin{equation}
  \label{eq:x-dgl-repara}
	\tilde{\dot{X}} = \frac{d \tilde{X}}{d \tilde{t}} =
	  \tilde{\omega} \frac {1 + \tilde{X}^2 - 2 \tilde{X} - \tilde{R}_z^2} {(1 - \tilde{X}) (1 - \tilde{X}^2 - \tilde{R}_z^2) } \,,
\end{equation}
where we have rescaled all lengths by the total contour length 
according to 
  $\tilde{R}_z = R_z / L$
and 
  $\tilde{X} = X/L$,
and introduced the characteristic 
time scale
  $\tilde{t}=t /(L^3 \zeta)$
and adsorption energy
  $\tilde{\omega}=\omega L$.
In deriving \eqref{eq:x-dgl-repara} we implicitly assume microscopic
relaxation processes such as molecular bending or stretching 
(as considered in ref.~\cite{Hanke05})
to equilibrate on much faster time scales than 
the global polymer geometry. The two poles in \eqref{eq:x-dgl-repara} correspond to the asymptotic limits
of complete adsorption
  ($X=L$) and
complete desorption
  ($L^2=X^2 + R_z^2$).
The force acting in the vertical direction, \ie the force being measured by the AFM, is
\begin{equation}
  \label{eq:vert-force}
	\frac{F_z}{\kBT} = \left(\pd {E}{R_z} \right)_{X} = \omega \frac {\tilde{R}_z} {1 - \tilde{X}} = \omega \frac {R_z} {R - R_x} = - \frac {R_z} {R_x} \frac{F_x}{\kBT} \,,
\end{equation}
where the latter relation is equivalent to the observation that a flexible string
can only support force along its contour.

\begin{figure}
  \includegraphics[height=70mm]{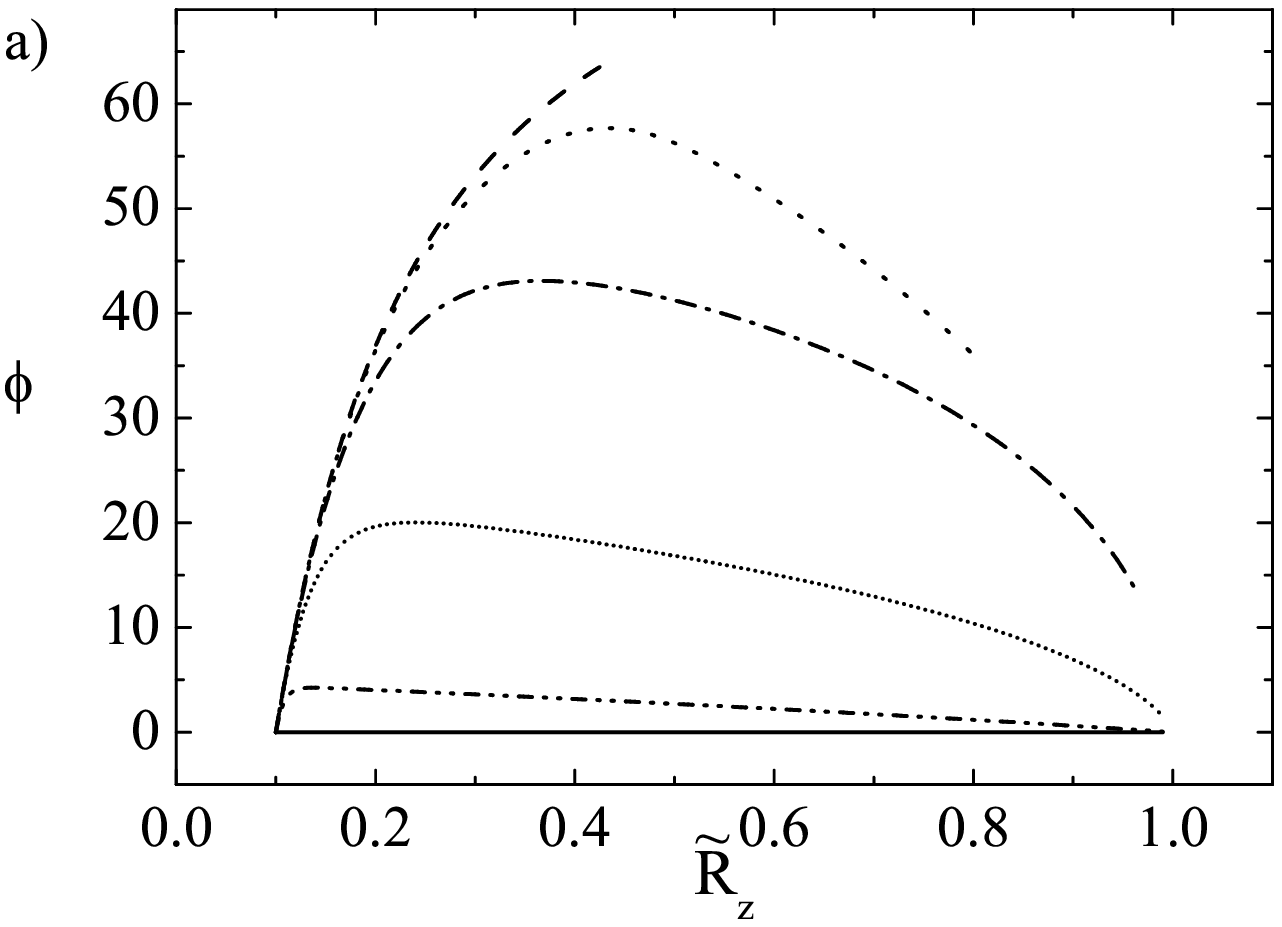}\\
  \includegraphics[height=70mm]{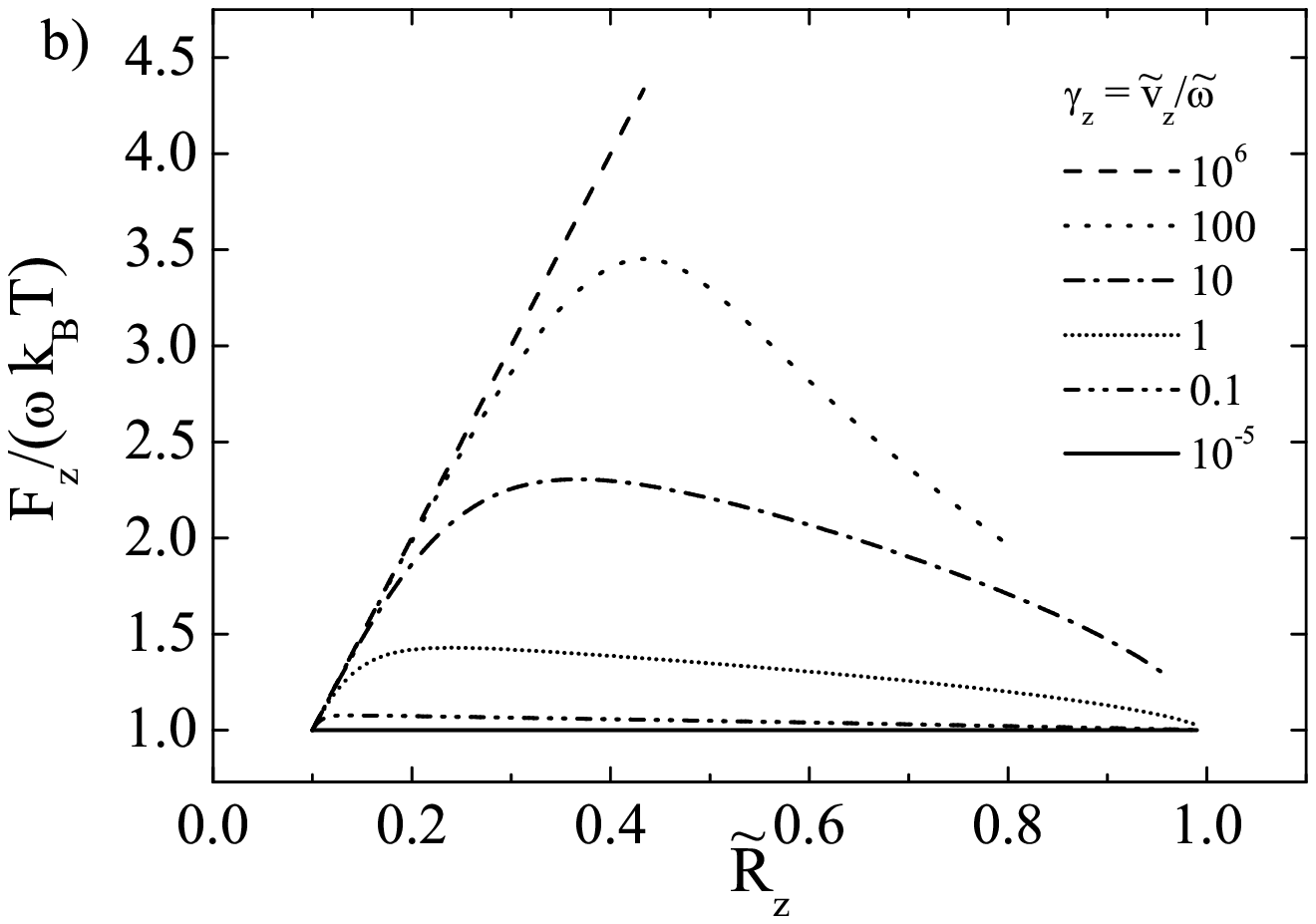}
  \caption {Vertical pulling:
Evolution of a)~the polymer angle $\phi$ and b)~the vertical force in units of the adsorption energy per unit length $F_z/(\omega\kBT)$ as functions of the normalized tip height $\tilde{R_z}=R_z / L$. 
Results are shown for different values of the rescaled 
friction coefficient $\gamma_z=v_{z} \zeta L / \omega$, 
including the limiting cases of pure slip ($\gamma_z = 10^{-5}$, solid line) 
and pure stick ($\gamma_z = 10^6$, broken line).}
	\label{fig:vert-pull}
\end{figure}

\paragraph{Vertical cantilever motion\\}
\label{sec:vertical-motion}
In a typical single molecule force spectroscopic experiment the AFM z-piezo-element is moved with a constant velocity. The measured desorption forces are in the $100\un{pN}$ to $\un{nN}$ range. With typical force constants of the cantilevers being $0.1\un{Nm^{-1}}$, the cantilever bending response is of the order of $10\un{nm}$. Since the vertical position of the cantilever is generally changed in the $100\un{nm}$ to $\um$ scale and thus much larger than the bending, we can consider 
the velocity of the cantilever tip to be constant,
\begin{equation}
  \label{eq:vert-distance}
	R_z = v_{z} t + R_{z0} \,,
\end{equation}
where $v_{z}$ is the vertical velocity of the tip. 
The differential equation representing the friction-adsorption force equilibrium, 
 \eqref{eq:x-dgl-repara}, is still valid if \eqref{eq:vert-distance} is inserted. 
It proves useful to rewrite the differential equation slightly,
\begin{equation}
  \td {\tilde{X}}{\tilde{R}_z} = \td {X}{t} \td {t}{R_z} = \frac {\dot{X}} {v_{z}} =
    \gamma_z^{-1} \frac {1 + \tilde{X}^2 - 2 \tilde{X} - \tilde{R}_z^2} {(1 - \tilde{X}) (1 - \tilde{X}^2 - \tilde{R}_z^2) } \,,
  \label{eq:x-dgl-repara-const-vert-veloc}
\end{equation}
where
  $\gamma_z=\tilde{v}_{z} / \tilde{\omega}= v_{z} \zeta L/\omega$
is the only material parameter remaining, measuring the ratio of friction versus adsorption strength.
The differential equation is solved using standard finite difference techniques \cite{Press92}. 
In \figref{fig:vert-pull} we show results for the polymer angle $\phi$ and for the vertical
force $F_z$ as a function of the cantilever height for different 
values of $\gamma_z$. In the very slippery case 
(solid line,
  $\gamma_z = 10^{-5}$) 
the desorbed polymer stays vertical (and thus 
  $\phi \simeq 0$)
since the polymer can freely glide over the surface; the force shows a perfect plateau.
In the very sticky case (upper broken line,
  $\gamma_z = 10^{6}$),
the force grows linearly with distance, as predicted by \eqref{eq:force-straightened-stick}.
In the intermediate case, the angle shows non-monotonic behavior:
initially, the constantly moving cantilever exerts a growing force on the polymer, but 
as more polymer is desorbed the friction force decreases.
Likewise, the vertical forces are much higher than the plateau force observed 
in the quasistatic limit,
which is due to a combination of friction (dissipative) and geometric effects.

\begin{figure}
  \includegraphics[height=70mm]{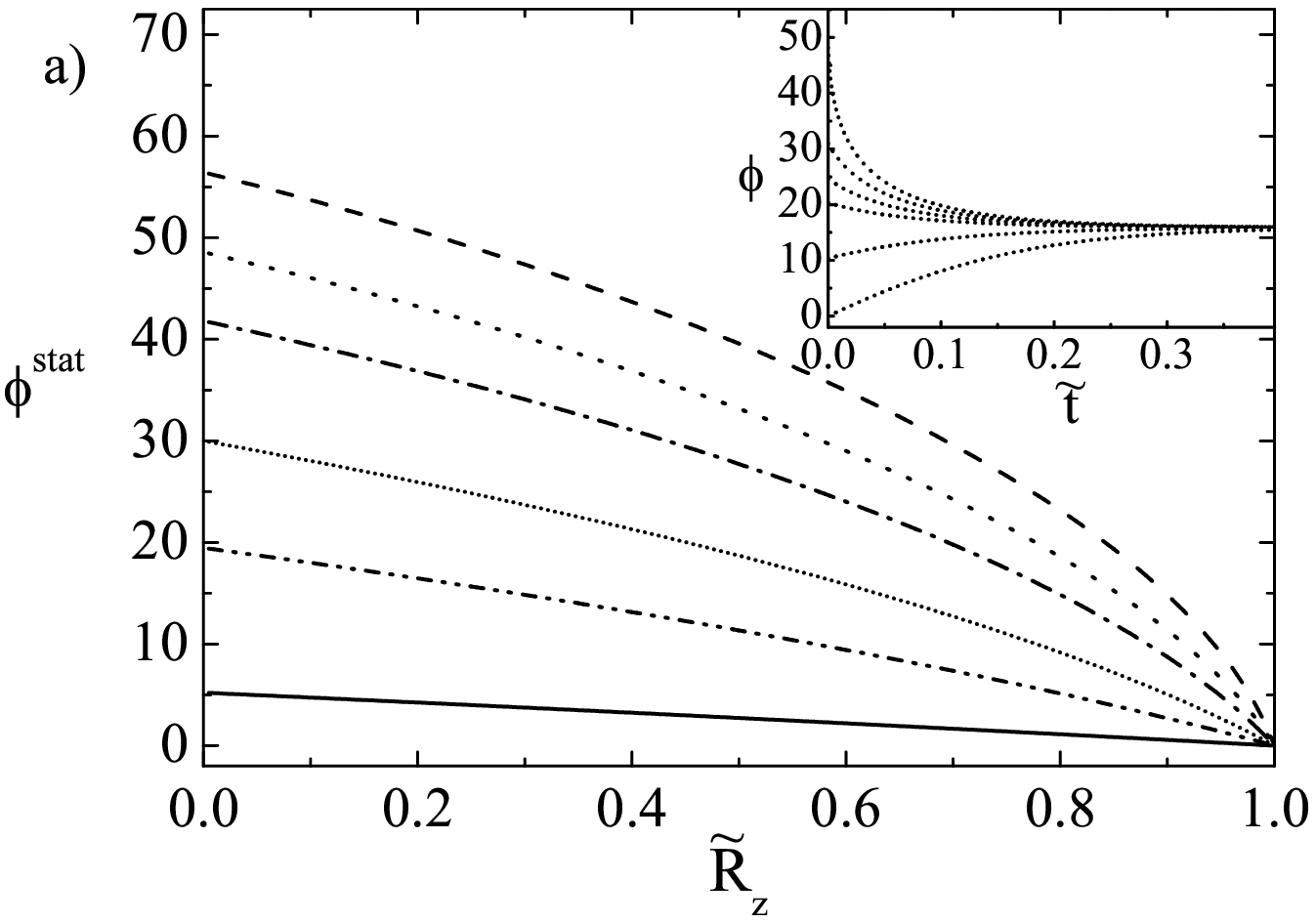}\\
  \includegraphics[height=70mm]{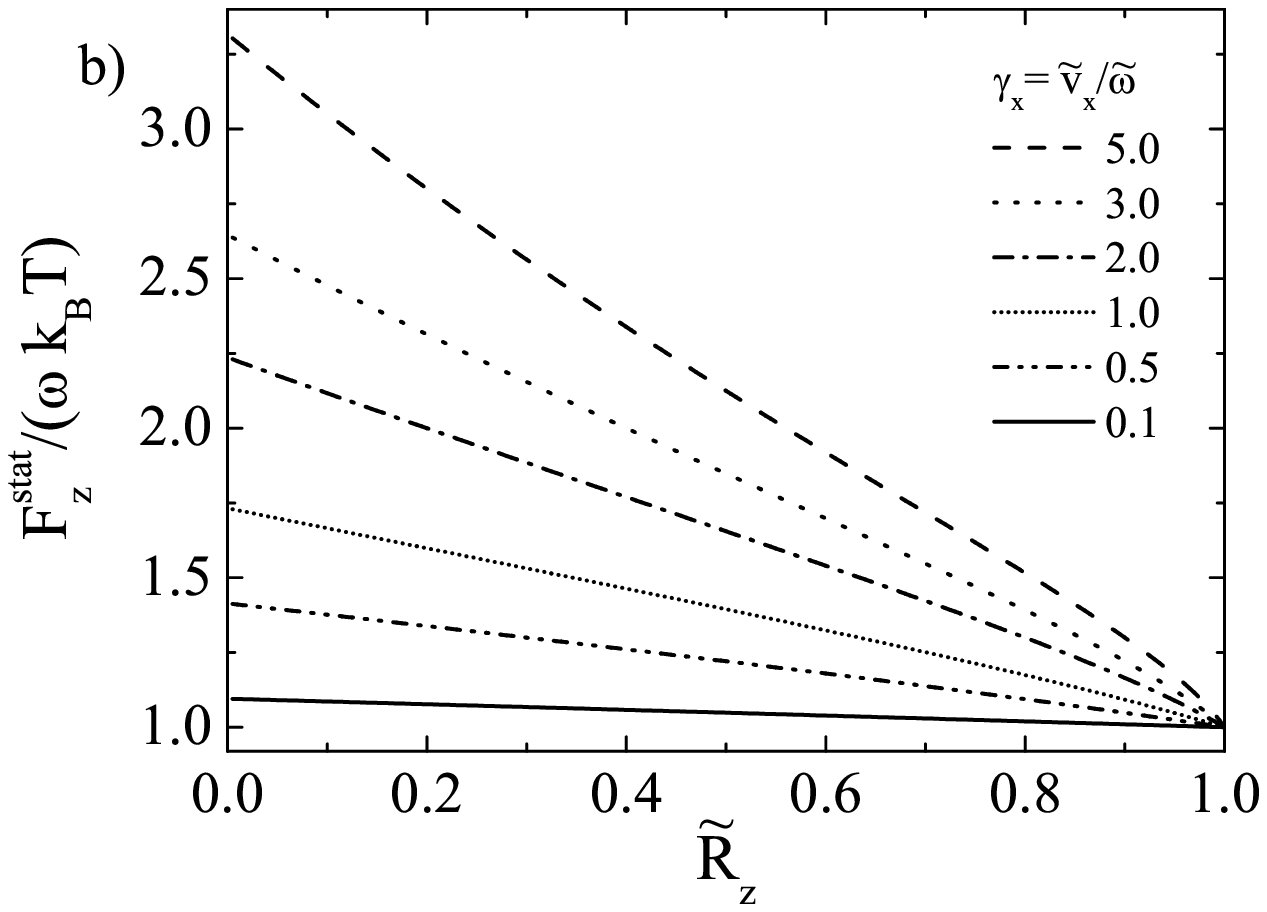}
  \caption {Results for horizontally moving cantilever.	
  a)~Stationary polymer angle $\phi^{stat}$ and b)~stationary vertical force 
  in units of the adsorption energy per unit length,
    $F_{z}^{stat}/(\omega\kBT)$ 
  as functions of the constant, normalized pulling height
    $\tilde{R_z}=R_z/L$. 
  Results are shown for a set of different values of the friction parameter
    $\gamma_x=v_{x} \zeta L / \omega$, 
  \ie for different pulling speeds $v_x$, polymer lengths $L$ or 
  sliding friction coefficients $\zeta$. 
  The inset shows the time evolution of the angle $\phi$ for fixed
  $\gamma_x = 1$ and $\tilde{R_z}=0.6$ for starting geometries with different angles
  as a function of rescaled time 
    $\tilde{t}=t L^3/\zeta$.}
	\label{fig:hori-pull}
\end{figure}

\paragraph{Horizontally moving tip\\}
\label{sec:horizontal-motion}
Friction effects are most clearly exhibited when a surface-adsorbed molecule
is moved laterally over the surface. In the presence of
a finite lateral cantilever velocity $v_x$, the differential equation
for the lateral polymer extension $X$ \eqref{eq:x-dgl-repara} is slightly modified
\begin{equation}
  \label{eq:x-dgl-hori-movement}
	\tilde{\dot{X}} = \tilde{\omega} \frac {1 + \tilde{X}^2 - 2 \tilde{X} - \tilde{R}_z^2}{(1 - \tilde{X})(1 - \tilde{X}^2 - \tilde{R}_z^2) } + \tilde{v}_{x} \,,
\end{equation}
where
  $\tilde{v}_{x}=v_{x}\zeta L^2$.
The stationary geometry for horizontal pulling is achieved for long enough pulling times
  $t \rightarrow \infty$
and can be derived from the above equation by setting
  $\tilde{\dot{X}}=0$.
For a given pulling height, $\tilde{R}_z$, this fully determines the
geometry of the adsorbed polymer and thus the force that is acting on the cantilever tip.
In \figref{fig:hori-pull} the angle and the (vertical) force acting on the cantilever in the stationary state ($\phi^{stat}$ and $F_z^{stat}$) are shown as functions of varying height $R_z$ for different friction to adsorption energy parameters $\gamma_x$. Both decrease with increasing pulling height because the alignment becomes more vertical, and increase with increasing friction parameter $\gamma_x$ in a monotonic but non-trivial way.
The time evolution of the angle for different starting geometries is shown in the inset. 
From measurements of the stationary force acting at different pulling velocities or heights, 
the frictional coefficient can be inferred, once $\omega$ has been determined in sufficiently 
slowly performed vertical desorption experiments carried out under the same conditions. 
$L$ can be determined in the same experiments as final height before complete desorption. By this procedure, 
measurements of frictional coefficients of single molecules on solid surface could be conducted, allowing 
for mapping out the dependence on parameters such as $\pH$ or added salt concentration. 
There is reason to believe that for many polymer-substrate combinations, the 
friction parameter is quite small, that is
  $\gamma_x=\tilde{v}_{x}/\tilde{\omega} \ll 1$. 
In a small-$\gamma_x $ expansion we obtain for the stationary vertical force
  $F_{z}^{stat} / (\omega \kBT ) = 1 + ( v_{x} L \zeta / \omega) (1 - R_z / L ) + \mathrm{O} (\gamma_x^2) $.
Similarly, the lateral force becomes
  $F_{x}^{stat} / (\omega \kBT ) = \gamma_x (1 - \tilde{R}_z ) + \mathrm{O} (\gamma_x^3) $.
These limiting laws will allow for straightforward fitting of experimental data.
Note that even in the case when the friction coefficient $\zeta$ is small,
the effective friction parameter
  $\gamma_x = v_{x} L \zeta / \omega$
can be made sufficiently large by choosing very long polymers, increasing the pulling velocity or by changing the solvent quality as suggested by recent experiments \cite{Haupt02}.
Thus there is hope that indeed AFM data can be used to extract friction coefficients 
of adsorbed single polymers, which is an important parameter in the context of adsorption
and desorption kinetics. What remains to be elucidated is the microscopic 
mechanism behind the friction of adsorbed polymers, \iec the dependence 
on the range of surface-monomer interaction,
the surface shape, and the distribution and density of interacting surface groups.
Note that in contrast to surface nanofriction studies with the AFM or SFA \cite{Krim02,He99}, in our case the normal force is self-adjusted by the surface-polymer attraction forces and not a free parameter.
We have an intimate coupling between the 
adsorption strength $\omega$ and the friction coefficient $\zeta$, and it is
the interplay embodied in the effective friction parameter $\gamma_{x,z}$,
which determines the resulting behavior. Interesting physics is expected as 
  $\omega b \rightarrow 1$
when polymer conformational fluctuations are modified due to the forced surface sliding; 
AFM methods, however, cannot probe this regime.
In addition, it will be interesting to compare results with measurements and theoretical 
studies of friction between small, uncharged adsorbates and metal substrates
under air atmosphere or vacuum \cite{Smith96,Persson99}.
These studies address the tribologically important aspect of 
surface contamination by exposure to air.

\paragraph{Acknowledgements}
We thank C.~Friedsam, H.~E.~Gaub, T.~Hugel, and F.~K\"{u}hner for fruitful discussions and the DFG (SFB~486) for financial support.

\newpage

\paragraph{Figures}{\ }
\vspace{5cm}

\newpage

\newpage

\end{document}